\documentclass[11pt]{article}
\usepackage{a4}
\usepackage[dvips]{epsfig}
\usepackage{latexsym}
\usepackage{amssymb}
\usepackage{graphicx}
\usepackage{psfrag}
\def\setb@se#1{\baselineskip=#1 \normalbaselineskip=#1}
\setb@se{14pt}
\def\setb@se#1{\baselineskip=#1 \normalbaselineskip=#1}
\setb@se{14pt}

\newcommand{\be}{\begin{equation}}
\newcommand{\ee}{\end{equation}}
\newcommand{\beqn}{\begin{eqnarray}}
\newcommand{\eeqn}{\end{eqnarray}}
\newcommand{\bsub}{\begin{subeqnarray}}
\newcommand{\esub}{\end{subeqnarray}}

\newcommand{\scr}{\scriptscriptstyle}

\def\bogo{{Bogomol'nyi}}

 \def\pa{\partial}
 \def\l{\left}
 \def\r{\right}

\newcounter{subequation}[equation]

\makeatletter \expandafter\let\expandafter\reset@font\csname
reset@font\endcsname
\newenvironment{subeqnarray}
  {\arraycolsep1pt
    \def\@eqnnum\stepcounter##1{\stepcounter{subequation}{\reset@font\rm
      (\theequation\alph{subequation})}}\eqnarray}%
  {\endeqnarray\stepcounter{equation}}
\makeatother

\begin{document}

\title{Superconducting vortices in Semilocal Models}
\author{P\'eter Forg\'acs$^{1,2}$, S\'{e}bastien Reuillon$^{1}$ and  Mikhail S.~Volkov$^{1}$\\
$^{1}$Laboratoire de Math\'{e}matiques et Physique Th\'{e}orique\\
CNRS-UMR 6083, Universit\'{e} de Tours\\
Parc de Grandmont 37200 Tours, France\\
$^{2}$MTA RMKI, H-1525 Budapest, P.O.Box 49, Hungary}

\maketitle
\begin{abstract}
It is shown that the SU(2) semilocal model --
the Abelian Higgs model with two complex scalars --
admits a new class of {\sl
stationary}, straight string solutions carrying a {\sl
persistent current} and having finite energy per unit length. In
the plane orthogonal to their direction they correspond to a
nontrivial deformation of the embedded Abrikosov-Nielsen-Olesen
(ANO) vortices by the current
flowing through them. The new
solutions bifurcate with the ANO vortices in the limit of
vanishing current.
They can be either static or stationary.
In the stationary case
the relative phase of the two scalars rotates at constant
velocity, giving rise to an electric field and angular momentum,
while the energy remains finite.
The new static vortex solutions have lower energy than the
ANO vortices and could be of considerable importance in various
physical systems (from condensed matter to cosmic strings).
\end{abstract}
\maketitle

It is difficult to overemphasize the importance of ``topological defects'' (monopoles,
vortex lines, domain walls, etc) resulting from the spontaneous breaking
of global or local symmetries in modern physics,
we refer to the some of the numerous reviews \cite{review}.
There is an interesting class of models
in which both global and local symmetries are simultaneously broken,
in the most economical way, with the minimal number of scalar fields.
These models, dubbed semilocal models, have some remarkable features,
see the detailed review \cite{semilocal-review}.
Perhaps the most remarkable point is that semilocal models exhibit vortices,
despite the first homotopy group of the vacuum manifold being trivial
\cite{vac-ach,hin}. The simplest example of such a model
is when electromagnetism is coupled to {\sl two}
charged complex scalars transforming as a doublet under a global
SU(2) symmetry (extended Abelian Higgs model).
This simple case is made quite interesting by noting
that it represents the bosonic sector of the Electroweak theory in the limit
that the Weinberg angle, $\theta_{\rm W}=\pi/2$ \cite{vac92}.
Furthermore such multicomponent Ginzburg-Landau models
are also used in condensed matter
theory, for example in the description of unconventional superconductors 
\cite{Sigrist}, in modeling neutral two-component plasmas \cite{Torokoff}.
The Abrikosov-Nielsen-Olesen (ANO) vortex solutions \cite{NO},
characterized by the mass ratio, $\beta=m_{\rm s}/m_{\rm v}$
($m_{\rm s}$ resp.\ $m_{\rm v}$
denoting the mass of the scalar resp.\ of the vector fields),
are embedded into this theory.
The parameter $\beta$ also distinguishes superconductors of type I ($\beta<1$)
from type II ($\beta>1$).
The case $\beta=1$ is quite special for here,
instead of there being a unique vortex solution (the ANO vortex),
there is a family of them labelled by a complex number of the same energy,
satisfying the \bogo\ equations \cite{bog}.
The stability of the embedded ANO vortices in Abelian models with an extended
Higgs sector was examined in
Ref.\ \cite{hin}, and it was shown there that they are stable only if
$\beta\leq1$. The counterparts of these solutions are the Z-strings
in the Standard model which have been shown to be stable for
$\sin^2\theta_{\rm W}\gtrsim0.9$ \cite{semilocal-review}.

In this Letter we present a new class of straight flux-tube solutions
(strings) of the SU(2) symmetric semilocal model which are translationally symmetric
along the $z=x_3$ direction.
The new solutions form a one parameter family for any fixed value of $\beta>1$.
The main feature distinguishing them from the previously known string or
equivalently vortex solutions in the $(x_1,x_2)$ plane, is due to
the presence of a relative phase between the two scalars,
depending linearly on $z$. Because of this relative phase (twist) there is also
a {\sl persistent current} flowing along the $z$-direction, so in fact these new
objects are not genuinely planar vortices.
The relative phase may also have a linear dependence on the time coordinate, $x_0$,
in which case the vortices are {\sl stationary}
and possess a nonzero momentum, angular momentum, as
well as an electric field. Due to the strong (exponential)
localization of the fields the vortices have finite energy per
unit length.
The energy of the current carrying solutions is {\sl lower} than that of the
corresponding ANO vortex, therefore they should have
better stability properties.
When the current vanishes the solution {\sl bifurcates} with the embedded ANO vortex.
In fact the well known ``magnetic spreading'' instability of the embedded type II ($\beta>1$)
ANO vortex in the semilocal model
can be understood as a signal of bifurcation with a new solution branch,
which carry a nonvanishing current.
The Z-string counterparts of
the new solutions should also be present in the Electroweak
model and they may have better stability properties because of the localizing
effect of the current.
Our solutions can also be interpreted as superconducting cosmic strings
of the bosonic type, but without an additional U(1) gauge field
as in the prototype U(1)$\times$U(1) model \cite{Witten}.
There are also some important differences between the superconducting strings
in Witten's model as compared to the semilocal ones.
For example the current carrying semilocal strings are strongly localized.

The (suitably rescaled) Lagrangian of the SU(2) semilocal theory can be written as
\be
\label{Lagrangian}
{\cal L} = \frac{1}{2g^2}\l\{-
\frac{1}{2}F_{\mu\nu}F^{\mu\nu}+|D_\mu\Phi|^2 -
\frac{\beta^2}{4}(|\Phi|^2 -1)^2\r\}
\ee
where  $\Phi=(\phi_1,\phi_2)$,
$D_\mu = \pa_\mu-iA_\mu$,
$|\Phi|^2 ={\bar \phi}_1\phi_1+{\bar \phi}_2\phi_2$,
and $g$ is the coupling constant, which will be set to $1$ for simplicity.
The breaking of the U(1) gauge symmetry leads to the
physical spectrum consisting of a vector particle with mass $m_{\rm v}=
g\eta$, two Nambu-Goldstone bosons, and a Higgs
scalar with mass $m_{\rm s} = \beta\eta$, $\eta$ is the vacuum expectation value
of the scalar fields.
The most general translationally symmetric (in the $x_3=z$ direction)
and stationary Ansatz can be written as
\beqn\label{ansatz1}
&A_\mu =  (A_{\scr 0}(x_1,x_2)\,,A_{\scr i}(x_1,x_2))\quad i=1,2,3\nonumber\\
&\phi_1 = f_1(x_1,x_2) \quad
\phi_2 =  f_2(x_1,x_2)e^{i(\omega_0 t+\omega_3 z)}\,,
\eeqn
where $(\omega_0,\omega_3)=\omega_\alpha$, are real parameters. The most important feature
of the Ansatz (\ref{ansatz1}) is that the change in the phase of
the scalar fields under an infinitesimal translation is compensated by a
nontrivial gauge transformation. In other words a space-time translation
moves the fields {\sl along gauge orbits}. Also the Ansatz (\ref{ansatz1})
reduces the global SU(2) symmetry to U(1).
The Noether current associated to the remaining global U(1) symmetry is then simply
$J_\mu=i({\bar \phi}_2D_\mu\phi_2-\phi_2\overline{D_\mu\phi}_2)/2$. Therefore
the solutions described by the Ansatz (\ref{ansatz1})
have a conserved Noether charge, ${\cal I}_0$, as well as a current,
${\cal I}_3$, flowing along the string where
\be
{\cal I}_\alpha=\pm\eta^2\int d^2x (\omega_\alpha-A_\alpha){\bar \phi}_2\phi_2\,,\quad\alpha=(0,3)\,.
\ee
The equations of motion imply the following ``Gauss-law'' constraints
on $A_\alpha=(A_0,A_3)$
in the $(x_1,x_2)$ plane:
\be\label{Gauss-law}
\triangle A_\alpha-A_\alpha |\Phi|^2+\omega_\alpha{\bar \phi}_2\phi_2=0\,.
\ee
Now exploiting the ``constraint'' equations (\ref{Gauss-law})
one can establish the following lower bound on the total energy per unit length:
\be\label{bound}
E\geq \pi |n|\eta^2+|\omega_\alpha{\cal I}_\alpha|\,,
\ee
where $n$ is an integer determining the magnetic flux.
Furthermore from Eqs.\ (\ref{Gauss-law}) it also
follows that for any regular solution $A_0=\omega_0 A_3/\omega_3$ ($\omega_3\ne0$),
i.e.\ $A_0$ is determined by $A_3$.
This is related to the fact that one can still perform Lorentz
transformations (boosts)
in the $(t,z)$ plane, which mix the $A_0$ and $A_3$ components of
the vector potential, and also transform the parameters $(\omega_0\,,\omega_3)$
of the Ansatz (\ref{ansatz1}).
In fact it is only their Lorentz invariant combination,
$\omega^2=\omega_3^2-\omega_0^2$, that appears in the equations of motion.
Therefore if $\omega^2>0$, by a Lorentz boost one can always achieve
$\omega_0=0\,, A_0=0$, i.e.\ it is sufficient to consider the {\sl static} case only.
It should be pointed out here that the general Ansatz (\ref{ansatz1})
has been also considered by Abraham \cite{Abraham} in the same model.
He considered, however, exclusively
the \bogo\ equations existing only in the very special case $\beta=1$.
As found in Ref.\ \cite{Abraham} this implies further $\omega_3=\omega_0$ i.e.\
$\omega^2=0$, which is referred to as the chiral case,
when due to the lack of localization
the energy of the solutions exhibited in Ref.\ \cite{Abraham} (with winding
number $n=1$) diverges.

To simplify further the problem of finding current carrying string solutions
of the semilocal theory Eq.\ (\ref{Lagrangian}),
we also impose rotational symmetry in the $(x_1,x_2)$ plane,
leading finally to the following Ansatz in polar coordinates:
 \beqn\label{ansatz2}
&A_0=a_0(r)\,,A_r=0\,, A_\varphi=na(r)\,,A_3=\omega a_3(r)\,,\nonumber\\
&\phi_1 = f_1(r)e^{in\varphi} \quad
\phi_2 = f_2(r)e^{im\varphi}e^{i(\omega_0 t+\omega_3 z)}\,,
\eeqn
where $m=0,\ldots n-1$.
In this Letter we shall concentrate on the simplest
$m=0$ case, a more exhaustive investigation will be published \cite{FRV}.
The static (i.e.\ $a_0=0$) field equations for $m=0$ become:
\bsub\label{stateqs}
\frac{1}{r}(ra_3')'&=&a_3|f|^2-f_2^2 \,,\ {\rm with}\ \;'\!=d/dr\,,\\
 r(\frac{a'}{r})'&=&a|f|^2-f_1^2 \,,\\
 \frac{1}{r}(rf_1')'&=&f_1\left[n^2\frac{(1\!-\!a)^2}{r^2}+\omega^2a_3^2-
\frac{\beta^2}{2}(1\!-\!|f|^2)\right],\\
 \frac{1}{r}(rf_2')'&=&f_2\left[n^2\frac{a^2}{r^2}+\omega^2(1\!-\!a_3)^2-
\frac{\beta^2}{2}(1\!-\!|f|^2)\right]\,.
\esub
Eqs.\ (\ref{stateqs}) admit a $4$-parameter family of local solutions regular at the origin
where $a=a^{(2)}r^2+O(r^4)$, $a_3=a_3^{(0)}+O(r^2)$,
$f_1=f_1^{(1)}r^n+O(r^{2+n})$, $f_2=f_2^{(0)}+O(r^2)$
where $a^{(2)}$, $a_3^{(0)}$, $f_1^{(1)}$, $f_2^{(0)}$
are free parameters.
For $r\to\infty$ the regular solutions decay exponentially:
$a=1+O\l(\exp{[-br]}\r)$, $a_3=O\l(\exp{[-br]}\r)$, $f_1=1+O\l(\exp{[-cr]}\r)$,
$f_2=O\l(\exp{[-\omega r]}\r)$, where the constants, $b$, $c$, determining the
exponential decay are given as
$b=\min(1,2\omega)$, $c=\min(2,\beta,2\omega)$,
and they also comprise a $4$-parameter family of local solutions.
From this asymptotic behaviour it is clear that $\omega>0$ is an essential parameter
to ensure exponential localization.
The reduced energy functional takes the form:
\beqn\label{energy}
E&=&\pi\eta^2\int_0^\infty rdr\left\{n^2\frac{a'^2}{r^2}+\omega^2a_3'^2 +|f'^2|+
n^2\left[\frac{(1-a)^2}{r^2}f_1^2+\frac{a^2}{r^2}f_2^2\right]+\right.\nonumber\\
& &\hspace{2.5cm}\left.\omega^2\l[ f_1^2a_3^2+f_2^2(1-a_3)^2\r]
+\frac{\beta^2}{4}\l(1-|f^2|\r)^2 \right\}\,,
\eeqn
from which it is apparent that the exponentially localized solutions have finite energy.
In view of the mathematical complexity of
Eqs.\ (\ref{stateqs}) we have resorted to numerical techniques to obtain regular
solutions.
Adapting a multishooting procedure
\cite{FoObReu}, we have succeeded in obtaining numerically families of regular solutions of
Eqs.\ (\ref{stateqs}), representing a new class of vortices; see Fig.\ 1
for a sample vortex profile.
\begin{figure}[t]
\hbox to\linewidth{\hss%
\psfrag{X}{$\beta=2\,,\;\;\omega=0.1$}
  \psfrag{x}{$\ln(1+r)$}
  \psfrag{f1}{$f_{\scr 1}$}
  \psfrag{f2}{$f_{\scr 2}$}
  \psfrag{a}{$a$}
  \psfrag{a3}{$a_{\scr 3}$}
  \resizebox{10cm}{7cm}{\includegraphics{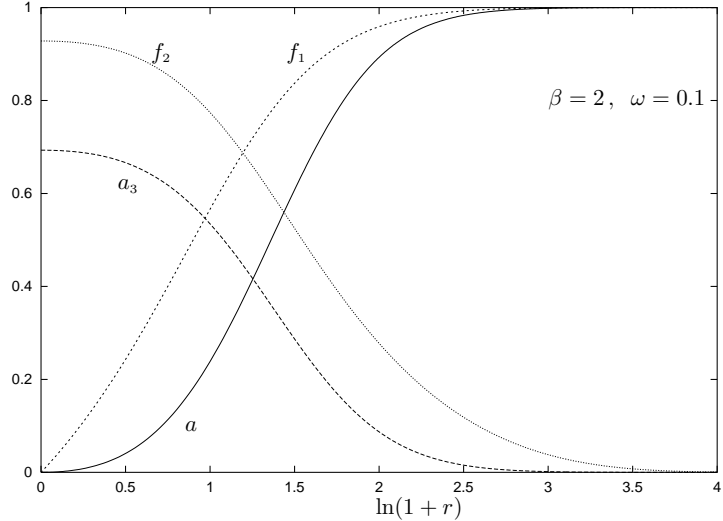}}%
    \hss}
\caption{\label{plotbackt}A superconducting vortex solution for $\beta=2,\omega=0.1$.}
\end{figure}
For a given value of $\beta>1$ our solution class is parameterized by $\omega$,
which varies in a finite interval $0<\omega\leq\omega_{\rm b}(\beta)$.
As the parameter $\omega$ tends to its maximal value, $\omega\to\omega_{\rm b}(\beta)$,
the functions $a_3\to0$, $f_2\to0$, while
$a\to a_{\scriptscriptstyle\rm ANO}$, $f_1\to f_{\scriptscriptstyle\rm ANO}$, i.e.\
the new class bifurcates with the corresponding ANO vortex.
This bifurcation can be understood by linearizing Eqs.\ (\ref{stateqs}) around
the ANO solution with $a_3\ll1$, $f_2\ll1$.
To first order $a_3\equiv0$ and the linearized $f_2$ amplitude
satisfies
\be\label{Schr}
 -\frac{1}{r}(rf_2')' +\left[\frac{a_{\scriptscriptstyle\rm ANO}^2}{r^2}
-\frac{\beta^2}{2}(1\!-\!f_{\scriptscriptstyle\rm ANO}^2)\right]f_2=-\omega_{\rm b}^2(\beta)f_2\,.
\ee
Eq.\ (\ref{Schr}) corresponds
precisely to the linearized stability problem of the embedded ANO vortex
in the semilocal theory
studied by Hindmarsh \cite{hin}. In agreement
with his results\footnote{Note that our $\beta^2$ corresponds to
$\beta$ of Ref.\ \cite{hin}.}, for any $\beta>1$  we find a single normalizable solution, $f_2$,
with negative eigenvalue, $-\omega_{\rm b}^2(\beta)$.
In Ref.\ \cite{hin} this bound state is a sign of instability of the semilocal
ANO vortex. In our case the very same bound state indicates
the bifurcation of a new branch  with that of
the ANO solution at $\omega=\omega_{\rm b}(\beta)$.
Quite importantly the energy of the new solutions (with $n=1$) is always
{\sl smaller} than that of the ANO vortex.
Therefore we expect the new solutions of winding number $n=1$ with a nonzero current
to be stable.

It is instructive to discuss also the $\beta=\infty$ limiting theory,
when the scalar fields are constrained $|f_1|^2+|f_2|^2\equiv1$,
(a gauged $\mathbf{CP}^1$-model) whose field equations can be written as:
\bsub\label{betainfeqs}
\frac{1}{r}(ra_3')'&=&a_3-\sin^2\theta,\;\
 r(\frac{a'}{r})'=a-\cos^2\theta,\\
\frac{1}{r}(r\theta')'&=&\frac{1}{2}\left[\omega^2(1-2a_3)-
n^2\frac{1\!-\!2a}{r^2}\right]\sin(2\theta)\,,
 \esub
where we have introduced $f_1=\cos\theta$, $f_2=\sin\theta$.
The profile functions of the current carrying vortex solution of Eqs.\ (\ref{betainfeqs})
are qualitatively similar to those of Eqs.\ (\ref{stateqs}) (comp.\ Fig.\ 1).
The energy of
the current carrying solutions is finite for any value of $\omega$,
whereas the energy of the $\beta=\infty$ ANO vortex diverges.
This difference is also reflected by the fact that $\omega_{\rm b}(\infty)=\infty$,
i.e.\ the parameter $\omega$ varies in the interval $(0,\infty)$.

The phase space of the new solution class is illustrated on
Fig.\ 2, choosing the total energy, $E$, the magnitude of the condensate
at the origin, $f_2^{(0)}$ and $\omega$ as parameters.
\begin{figure}[t]
\hbox to\linewidth{\hss%
    \psfrag{q}{\hspace{1cm}$f_{2}^{\scr (0)}$}
    \psfrag{E}{$E/\pi\eta^2$}
    \psfrag{sigma}{$\omega$}
    \psfrag{b1}{$\beta=1$}
    \psfrag{b1.5}{$\beta=\sqrt{2}$}
    \psfrag{b2}{$\beta=2$}
    \psfrag{b3}{$\beta=3$}
    \psfrag{binf}{$\beta=\infty$}
    \psfrag{Eno}{$E_{\scriptscriptstyle ANO}$}
    \resizebox{15cm}{10cm}{\includegraphics{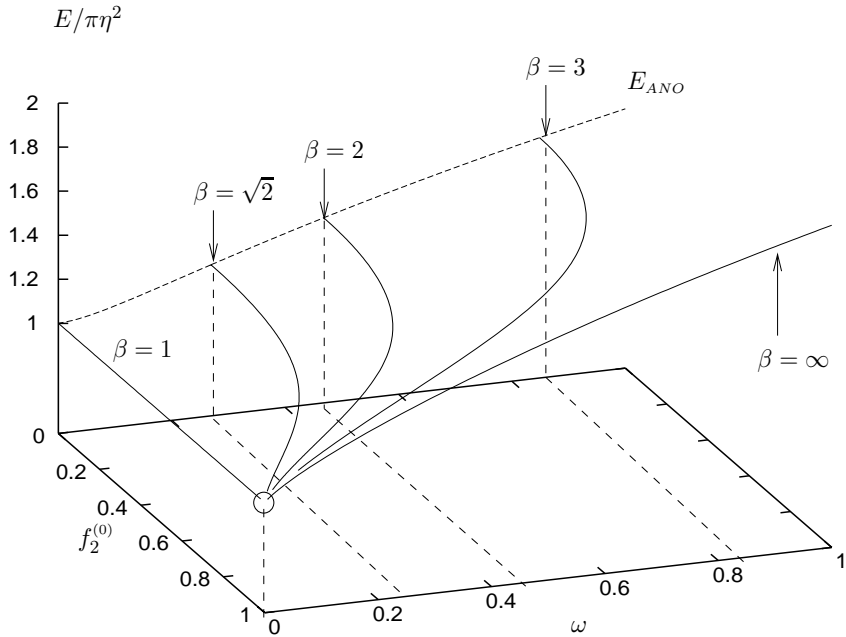}}%
    \hss}
    \caption{Phase space of the $n=1$ superconducting vortices.}
\end{figure}
On Fig.\ 2 a family of solutions for a fixed value of $\beta$ corresponds to a curve in the
($f_2^{(0)}$, $\omega$) plane, and we have indicated the corresponding ANO vortex,
located at the endpoint of the curve in the $f_2^{(0)}=0$ plane where
the new family bifurcates with it. The family of solutions for $\beta=\infty$ is represented
by the curve in the $f_2^{(0)}=1$ plane.
As the value of $\omega$ decreases
from $\omega_{\rm b}(\beta)$ towards $\omega=0$,
the energy of all superconducting solutions decreases towards
the \bogo\ bound, $E=\pi\eta^2$, {\sl without ever attaining it},
irrespectively of the value of $\beta$.
An important point is that solutions with nonzero values of $\omega$ do not have a smooth
$\omega\to0$ limit, i.e.\ the point $(f_2^{(0)}=1\,,\omega=0)$ does not belong
to the phase space as it is indicated on Fig.\ 2 by a circle.
In the $\beta=\infty$ theory one can obtain the following virial relation
\be\label{virial}
\int_0^\infty rdr\,\l\{n^2\frac{a'^2}{r^2}-\omega^2
[{a'}^2_3 +(1-2a_3)\sin^2\theta]\r\}=0\,,
\ee
which immediately implies that for $\omega=0$ the $\beta=\infty$ theory
{\sl does not} admit nontrivial finite energy solutions.
For finite values of $\beta$, a suitable change of scale in
Eqs.\ (\ref{stateqs}) shows that the effective
value of the scalar self coupling
is given by $\beta_{\rm eff}=\beta/\omega$. It follows that
the limit $\omega\to0$ also implies
$\beta_{\rm eff}\to\infty$, enforcing $|f_1|^2+|f_2|^2\to1$.
This explains the universal behaviour of the solutions
in the $\omega\to0$ limit.
A more detailed analysis shows \cite{FRV} that all solutions converge {\sl pointwise} to a
limiting configuration corresponding the $\mathbf{CP}^1$ lump analyzed in Ref.\ \cite{hin}.

In the limit $\omega\to0$ the value of the current, ${\cal I}_3$, is getting large
and it even seems to diverge
while $ \omega {\cal I}_3\to0$.
This behaviour of the current is in sharp contrast with
what is found in the U(1)$\times$U(1) model, where the current has a maximal value
above which the superconducting string ``goes normal'' \cite{current}.
The $z$-component of the current, $\tilde{\cal I}_3={\cal I}_3/(2\pi\eta^2)$, is depicted on
Fig.\ 3 as a function of the parameter $\omega$.
\begin{figure}[t]
\hbox to\linewidth{\hss%
  \psfrag{x}{\LARGE $\omega$}
  \psfrag{X}{\Large $\beta=\sqrt{2}$}
  \psfrag{b2}{\Large $\beta=2$}
  \psfrag{b3}{\Large $\beta=3$}
  \psfrag{binf}{\Large $\beta=\infty$}
  \psfrag{n=1}{\hspace*{-0.7cm}\Large $n=1$}
  \psfrag{n=2}{\Large $n=2$}

    \resizebox{8cm}{6cm}{\includegraphics{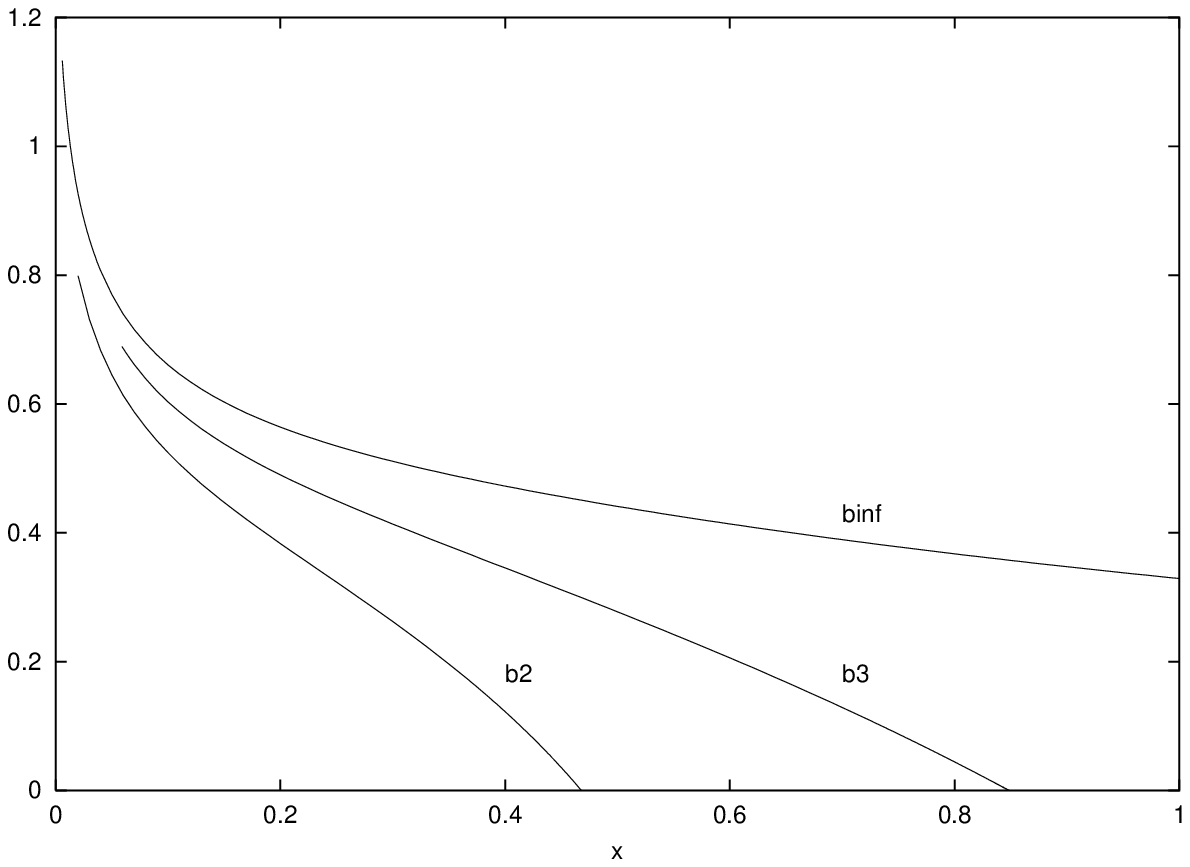}}%
\hspace{2mm}
    \resizebox{8cm}{6cm}{\includegraphics{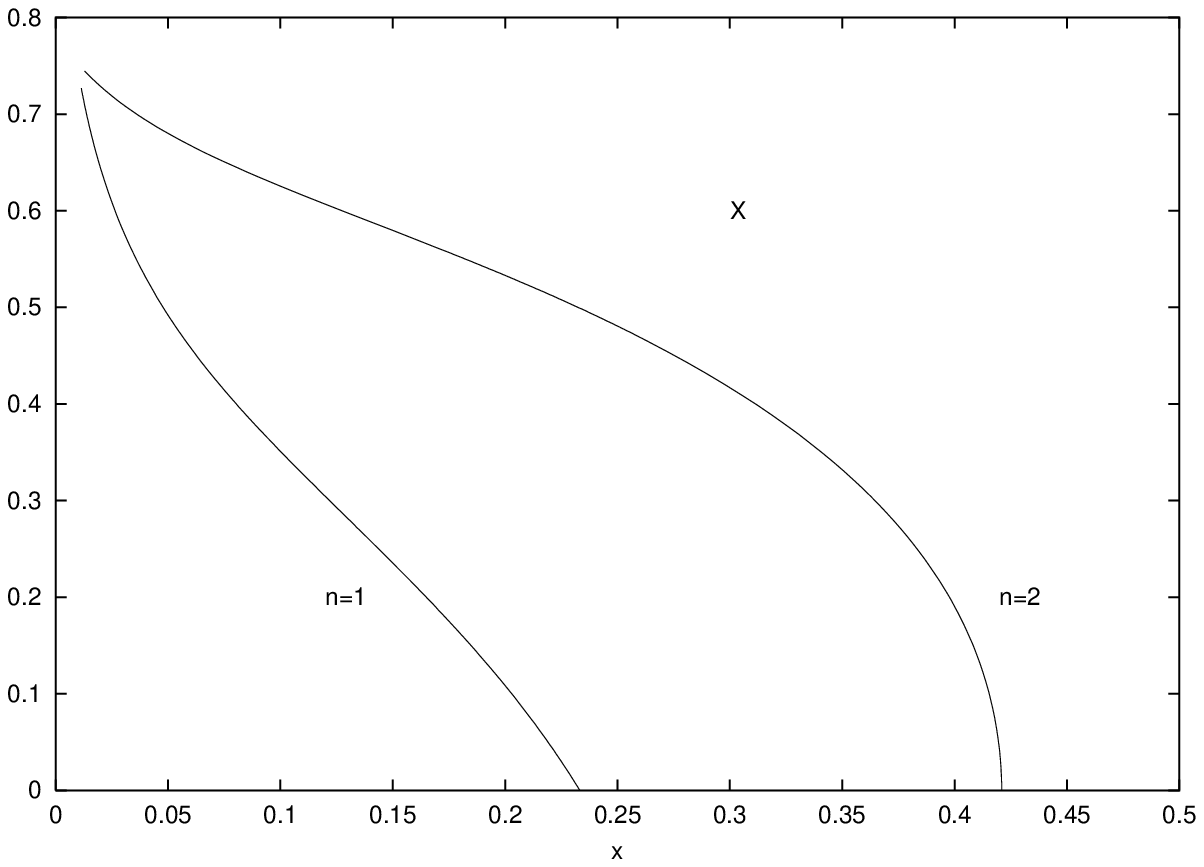}}%
\hss}
 \vspace{-1mm}
\caption{The current, $\tilde{\cal I}_{\scr 3}$ as a function of $\omega$
for $n=1$, $\beta=\sqrt{2},\,2,\,3,\,\infty$. (The curves terminate at the smallest
value of $\omega$ we could compute)}
\end{figure}

Since our vortex solutions exist only in type II superconductors
($\beta>1$) it is natural to ask as to their interaction energy
(which is well known to be repulsive for the type II ANO
vortices). Since we have superimposed vortex solutions (for $n>1$)
we can ask if it is energetically favourable for them to break up
into widely separated constituents.
To this end we depict
$E(n,{\cal I}_3)/(\pi n\eta^2)$, the energy per winding number of the
superimposed $n=2$ and $n=3$ solutions as a function of the
current flowing through them on Fig.\ 4. The graphs on Fig.\ 4
indicate that the interaction is repulsive up to a certain value
of $\tilde{\cal I}_3$, however, at least for the $n=2$ vortex it
becomes attractive when  $\tilde{\cal I}_3>0.71$. This suggest
that the $n=2$ superimposed vortex for large enough currents
becomes stable with respect to breakup into $n=1$ vortices unlike
the type II ANO vortices. If the interaction for separated
vortices has also an attractive phase this could have important
physical consequences (e.g.\ the attraction may significatively
change the intercommutation of colliding strings). Let us mention here
that in a somewhat different two-component Ginzburg-Landau model
similar non-monotonic behaviour of $E/n$ has been reported in
\cite{Babaev}.
\begin{figure}[t]
\hbox to\linewidth{\hss%
 \psfrag{y}{\large $E/n$}
  \psfrag{x}{\large $\tilde{\cal I}_3$}
  \psfrag{n1}{\large $n=1$}
  \psfrag{n2}{\large $n=2$}
  \psfrag{n3}{\large $n=3$}
 \resizebox{10cm}{7cm}{\includegraphics{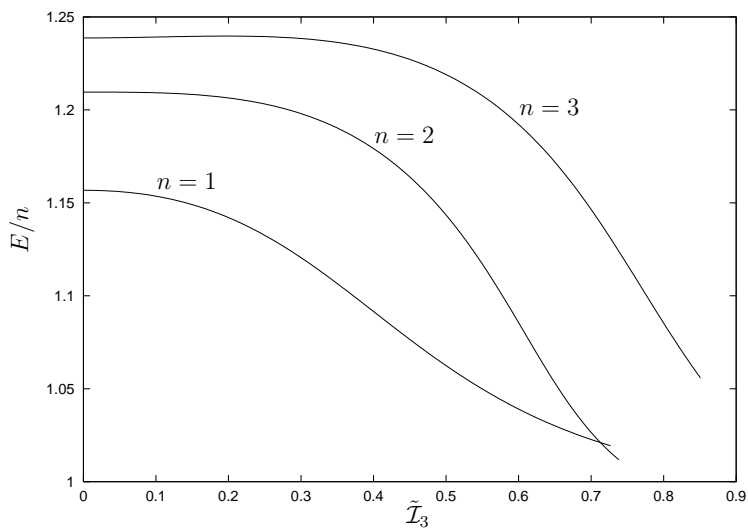}}%
\hss}
\vspace{-1mm}
\caption{\label{plotbackd}Energy/winding, $E/(\pi n\eta^2)$, of the $n=2,3$ vortices for
$\beta=\sqrt{2}$ in function of the current, $\tilde{\cal I}_3$.}
\end{figure}

Stationary solutions (with $\omega_0\ne0$) are obtained by a Lorentz boost
of the static ones along the $z$-axis.
The stationary vortices obtained this way have momentum, $P=\omega_0{\cal I}_3/\omega_3$,
angular momentum, $J=-\omega_0 n{\cal I}_3/\omega_3$, and they are surrounded by a radial
electric field $E_r=\omega_0 a'_3$. It is interesting that in our case the
electric field is screened, and the total energy of the superconducting vortices stays finite.

In conclusion such current carrying strings provide a new interesting class
of defects which may be realized in realistic physical systems described
by semilocal models and they may equally
find their applications in a cosmological context as cosmic strings
and may even be present in the Standard model of Electroweak interactions.
Finally we thank G.~Volovik for pointing out that twisted vortices
have also been investigated and even experimentally observed in superfluid
$^3$He \cite{Salomaa}.

\end{document}